\documentclass[a4paper,10pt]{article}
\usepackage{a4wide,authblk}

\usepackage{amssymb,amsmath,amsthm}
\usepackage{graphicx,rotating,color}
\usepackage{textcomp}
\usepackage{amstext}
\usepackage{amsfonts}
\usepackage{amsxtra}
\usepackage{hyperref}

\newcommand{\de}{\partial}
\newcommand{\R}{\mathbb R}

\newcommand{\vc}[1]{\boldsymbol{#1}}
\newcommand{\vt}[1]{\mathsf{#1}}
\newcommand{\tsp}{\mathsf{T}}

\newcommand{\str}{\varSigma}
\newcommand{\ad}{\operatorname{\mathrm{ad}}}

\title{Simulation of viscoelastic Cosserat rods based on the geometrically exact dynamics of special Euclidean strands}

\author{G.\ G.\ Giusteri$^{1,*}$, E.\ Miglio$^2$, N.\ Parolini$^2$, M.\ Penati$^2$, R.\ Zambetti$^{2,3}$}
\affil{$^1$Dipartimento di Matematica, Universit\`a degli Studi di Padova, Via Trieste 63, 35121, Padova, Italy\\
$^2$MOX, Dipartimento di Matematica, Politecnico di Milano, Piazza Leonardo da Vinci 32, 20133, Milano, Italy\\
$^3$ Tenaris (\texttt{www.tenaris.com})\\
$^*$E-mail for correspondence: giulio.giusteri@math.unipd.it}
\date{{\small \today}}

\begin{document}

\maketitle

\begin{abstract}
We propose a method for the description and simulation of the nonlinear dynamics of slender structures modeled as Cosserat rods. It is based on interpreting the strains and the generalized velocities of the cross sections as basic variables and elements of the special Euclidean algebra.
This perspective emerges naturally from the evolution equations for strands, that are one-dimensional submanifolds, of the special Euclidean group.
The discretization of the corresponding equations for the three-dimensional motion of a Cosserat rod is performed, in space, by using a staggered grid. The time evolution is then approximated with a semi-implicit method.
Within this approach we can easily include dissipative effects due to both the action of external forces and the presence of internal mechanical dissipation.
The comparison with results obtained with different schemes shows the effectiveness of the proposed method, which is able to provide very good predictions of nonlinear dynamical effects and shows competitive computation times also as an energy-minimizing method to treat static problems.
\end{abstract}

\section{Introduction}

The modeling and simulation of beams is of great importance in the engineering practice to analyze the configurations and stress distributions of a wide variety of mechanical structures, with sizes ranging from those of pipelines and cables to those of microactuators. When these structures are sufficiently slender or very flexible, they can undergo large displacements even in the small-strain and linear-response regime, and geometric nonlinearities must be taken into account to capture their mechanical behavior. 

Since the seminal work by Simo and Vu-Quoc~\cite{SimVuQ86}, the number of publications and numerical methods related to geometrically-exact beam models has been growing significantly.
Nevertheless, given the variety of applications and the different features pertaining to each method, no universal standard is available for an efficient simulation of such models.
On the other hand, it has become clear that the theory of special Cosserat rods (as presented for instance by Antman~\cite{Antman2005}) provides the optimal mathematical framework to deal with slender structures, as it comprises all of the classical beam models as special cases.

In the literature, we can find approximation schemes based on a discrete mechanical analogue for the rod, such as those by Bertails \emph{et al.}~\cite{BerAud06}, Bergou \emph{et al.}~\cite{BerWar08}, Giusteri \& Fried~\cite{GiuFri18}, Jung \emph{et al.}~\cite{JunLey11}, Lang, Linn \& Arnold~\cite{LanLin11}, and Linn~\cite{Lin20}.
A similar structure is shared by the finite-element approaches by Borri and Bottasso~\cite{BorBot94}, Cao, Liu \& Wang~\cite{CaoLiu06}, and Spillmann \& Teschner~\cite{SpiTes07}.
Several other methods are based upon discretizing the evolution equations for the continuum rod.
Many represent the rod via the position and orientation of nodal cross sections.
In this way, the computation of the strains and stresses associated with twist, bending, stretching, and shearing of the rod relies on interpolation procedures that introduce some important arbitrariness in the calculations~\cite{BauHan14}.
In other cases, nonlinear shape functions are used to approximate the configuration of the rod, as done by Patil \& Althoff~\cite{PatAlt11} and Howcroft \emph{et al.}~\cite{HowCoo18}.

There are approaches in which translational and rotational degrees of freedom are considered separately, as in the works by Simo \& Vu-Quoc~\cite{SimVuQ86}, Ibrahimbegovi\'c~\cite{Ibr95}, Betsch \& Steinmann~\cite{BetSte02}, Meier, Popp \& Wall~\cite{MeiPop14,MeiPop15}, Ga\'ce\v{s}a \& Jeleni\'c~\cite{GacJel15}, Bauer \emph{et al.}~\cite{BauBre16}, Yilmaz \& Omurtag~\cite{YilOmu16}, and Zupan \& Zupan~\cite{ZupZup16}.
Other methods consider the fundamental role of the special Euclidean group $\mathit{SE}(3)$ and the associated algebra $\mathfrak{se}(3)$. 
Sanders~\cite{San10}, Chirikjian~\cite{Chi10}, and Sonneville, Cardona \& Br\"uls~\cite{SonCarBru14,SonCar14} discretize the degrees of freedom at the group level, applying suitable techniques for the dynamics on a Lie manifold, while Zupan \& Saje~\cite{ZupSaj03,ZupSaj06}, \v{C}e\v{s}arek, Saje \& Zupan~\cite{CesSaj13},  Su \& Cesnik~\cite{SuCes11}, and Schr\"oppel \& Wackerfu{\ss}~\cite{SchWac16} perform the discretization on elements of the Lie algebra that are precisely the generalized strains of rod theory.

To simplify the derivation and the structure of the rod equations, a fundamental step is to view not only the strains but also the generalized velocities of the rigid cross sections as elements of the Lie algebra associated with the special Euclidean group of rigid body motions.
This perspective led Simo, Marsden \& Krishnaprasad~\cite{SimMar88} and Hodges~\cite{Hod90,Hod03} to derive the intrinsic rod equations from the variations of a Hamiltonian functional expressed solely in terms of Lie algebraic quantities.
Casting the equations in a linear space such as the algebra $\mathfrak{se}(3)$ has several computational advantages in reference to interpolation strategies and the imposition of linear constraints.

Starting from the approach of Holm \& Ivanov~\cite{HolIva14}, in Section~\ref{sec:equations} we derive, in a rather straightforward way, evolution equations that correspond to the mixed formulation proposed by Hodges~\cite{Hod90} with the addition of dissipative forces (both due to internal and external viscous phenomena) and of an equation that translates an important compatibility condition on the evolution of velocities and strains, necessary to close the system of partial differential equations.
Positional and rotational degrees of freedom never appear in the equations, since they are merely recovered following the evolution of the rod placement from the initial configuration as driven by the generalized velocities.
We then introduce a finite-difference scheme and discretize the evolution equations on a staggered grid so as to avoid shear-locking effects. In Section~\ref{sec:examples}, we show the effectiveness of our method by applying it to the solution of both static and dynamic problems that involve viscoelastic rods featuring possibly curved relaxed shapes and anisotropic cross sections.

We believe that the method presented here provides a synthesis of many of the features towards which the recent literature on geometrically exact rods is converging.
In particular, the theoretical setting enjoys a significant degree of mathematical transparency, the evolution takes place in a linear space with 
degrees of freedom represented in the most economic way, there are no limitations on the geometry of the cross sections and on the relaxed shapes, and the local nature of the representation allows for a straightforward application of external forces and the combination, by means of boundary conditions, of several structural elements.
Other intersting aspects, relevant for specific applications, are mentioned in Section~\ref{sec:conclusion}.

The target application that we have in mind is the study of the nonlinear dynamics of viscoelastic beams, but a strongly dissipative evolution can be used also as an alternative energy-minimization method to retrieve static solutions.
To assess the usefulness of our method, we implemented it in the Python language and compared its performance with a selection of published results, obtained with rather different schemes, and with results produced by an established commercial software.
We find that, in spite of the simplicity of our formulation and of the discretization schemes that we have adopted, the method achieves very good results in solving both static and nonlinear dynamical problems, with competitive computational times.

\section{The computational model}\label{sec:equations}

We propose a method for the description of the nonlinear dynamics of slender beams that is based on extensions of the $\mathit{SE}(3)$-strand equations described by Holm and Ivanov~\cite{HolIva14}, with a suitable mechanical interpretation of stresses and momenta as dual to strains and velocities.
Within this approach, we can easily include dissipative effects due to both the action of external forces and the presence of internal mechanical dissipation. 
Moreover, the relaxed shape of the rod can be arbitrarily prescribed.

\subsection{Rod kinematics}

We view a rod as a one-parameter family of rigid cross sections labelled by $s\in[0,L]$, where $L$ is a reference length.
Each cross section is characterized by the position $\vc x$ of its center of mass and three orthonormal vectors $\vc d_1$, $\vc d_2$ (in the cross-sectional plane) and $\vc d_3$ (orthogonal to the cross section).
The Euclidean transformation that translates the origin into $\vc x$ and rotates the Euclidean reference basis $(\vc e_1,\vc e_2,\vc e_3)$ into $(\vc d_1,\vc d_2,\vc d_3)$ is an element of the special Euclidean group $\mathit{SE}(3)$ of rigid body motions.
In this way, we can identify the rod at each time instant with an $\mathit{SE}(3)$-strand, that is a curve on $\mathit{SE}(3)$ parametrized by $s$.

We take as time parameter $t\in[0,+\infty)$ and define the field $\mathcal P:[0,L]\times[0,+\infty)\to\mathrm{Mat}_{4\times 3}(\R)$ by $\mathcal P:=(\vc x,\vc d_3,\vc d_1,\vc d_2)^\tsp$, namely, each $4 \times 3$ matrix $\mathcal P(s)$ has in its rows the components of the vectors $\vc{x}$, $\vc{d}_3$, $\vc{d}_1$, and $\vc{d}_2$ at $s$. Then we define
\begin{equation*}
\vt L(s,t):=
\begin{pmatrix}
0 & \sigma_3(s,t) & \sigma_1(s,t) & \sigma_2(s,t) \\
0 & 0 & -u_2(s,t) & u_1(s,t) \\
0 & u_2(s,t) & 0 & -u_3(s,t) \\
0 & -u_1(s,t) & u_3(s,t) & 0 
\end{pmatrix}\,.
\end{equation*}
The twist density $u_3$, curvatures $u_1$, $u_2$, the stretching $\sigma_3$, and the shearing densities $\sigma_1$ and $\sigma_2$ are the six strains that describe the shape of the rod.
We will identify the collection of them with the six-component strain vector $U=(u_3,u_1,u_2,\sigma_3,\sigma_1,\sigma_2)$.
The natural (relaxed) shape of the rod is given by fixing preferred strains $\bar{U}(s)$ for each value of $s$.
The placement of the rod in the three-dimensional ambient space is given by the solution of the ODE  
\begin{equation}\label{eq:Pprime}
\mathcal P'=\vt L\mathcal P,
\end{equation}
where the prime denotes differentiation with respect to $s$,
with the conditions $\mathcal P_0(t)$ given at $s=0$.
The nonlinear relation between the strains and the placement can be seen from the definitions
\[
\sigma_i:=\vc x'\cdot\vc d_i, \quad\text{for }i=1,2,3,\quad
u_1:=\vc d'_3\cdot\vc d_2,
\quad
u_2:=\vc d'_1\cdot\vc d_3,
\quad\text{and}\quad
u_3:=\vc d'_2\cdot\vc d_1.
\]
One of the advantages of our approach is to avoid using these relations in the simulation process.

\begin{figure}
\centering
\includegraphics[width=0.5\textwidth]{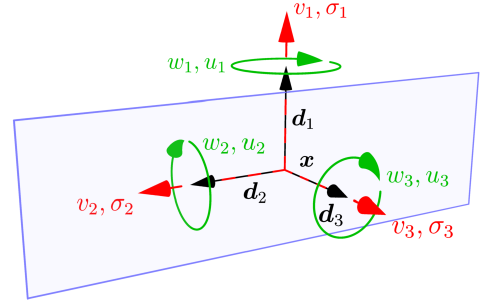}
\caption{The pose in space of each cross section is determined by the position $\vc x$ of its center of mass and three orthonormal vectors $\vc d_1$, $\vc d_2$ (in the cross-sectional plane) and $\vc d_3$ (orthogonal to the cross section). 
Generalized velocities and strains represent infinitesimal transformations applied to the pose in order to generate the movement in time (velocities) or the spatial transformation from one section to the other (strains). 
The fields $v_i$ and $\sigma_i$, for $i=1,2,3$, represent the intensity of infinitesimal translations of $\vc x$ in the direction $\vc d_i$ (red dashed arrows), while the fields $w_i$ and $u_i$ represent the intensity of infinitesimal rotations of the cross section about $\vc d_i$ (green solid arrows).}\label{fig:dofs}
\end{figure}

If we now consider the motion of each cross section we find that, for any given $s$, the time derivative of $\mathcal P$, denoted by a superimposed dot, is given by
\begin{equation}\label{eq:Pdot}
\dot{\mathcal P}=\vt\Omega\mathcal P,
\end{equation}
where the spin--velocity matrix $\vt\Omega$ has the very same structure of $\vt L$, with
\begin{equation*}
\vt\Omega(s,t):=
\begin{pmatrix}
0 & v_3(s,t) & v_1(s,t) & v_2(s,t) \\
0 & 0 & -w_2(s,t) & w_1(s,t) \\
0 & w_2(s,t) & 0 & -w_3(s,t) \\
0 & -w_1(s,t) & w_3(s,t) & 0 
\end{pmatrix}\,.
\end{equation*}
and features three linear velocities $v_i$ and three spins $w_i$ ($i=1,2,3$).
We will identify the collection of them with the generalized velocity vector $V=(w_3,w_1,w_2,v_3,v_1,v_2)$.
The definitions of velocities and spins in terms of the placement components read
\[
v_i:=\vc{\dot{x}}\cdot\vc d_i, \quad\text{for }i=1,2,3,\quad
w_1:=\vc{\dot{d}}_3\cdot\vc d_2,
\quad
w_2:=\vc{\dot{d}}_1\cdot\vc d_3,
\quad\text{and}\quad
w_3:=\vc{\dot{d}}_2\cdot\vc d_1.
\]
Figure~\ref{fig:dofs} summarizes the geometric and kinematic quantities involved in the rod description.

\subsection{Constitutive assumptions and Euler--Poincar\'e equations}

We can now introduce the momentum and stress fields,  $P$ and $\str$, as
\begin{equation}\label{eq:conjugate_vars}
P:=\mathbb M V\qquad\text{and}\qquad\str:=\mathbb A (U-\bar{U}),
\end{equation}
where the symmetric positive definite matrices $\mathbb M(s)\in\mathrm{Mat}_{6\times 6}(\R) $ and $\mathbb A(s)\in\mathrm{Mat}_{6\times 6}(\R)$ represent, respectively, the rigid-body inertia (linear density, determined by the geometry of the cross sections) and the elastic stiffnesses at each cross section.
The elastic response of the rod is here modeled with a linear function of the difference between the current strains and the preferred ones.

For the test cases considered in what follows, the rigid cross sections with surface area $A$ are assumed to have uniform mass density. We choose $\vc d_1$, $\vc d_2$, and $\vc  d_3$ aligned with the principal axes of inertia of each cross section and denote by $I_1$, $I_2$, and $I_3$ the corresponding second area moments.
We further assume that the elastic response does not couple different strain components.
Under these assumptions, the inertia and stiffness matrices take the form
\[
\mathbb M=\mathrm{diag}(\rho I_3/A,\rho I_1/A,\rho I_2/A,\rho,\rho,\rho)\qquad\text{and}\qquad \mathbb A=\mathrm{diag}(GI_3,EI_1,EI_2,EA,GA,GA),
\]
where $E$ is the Young modulus of the material and $G$ is the shear modulus, related to $E$ by $G=E/(2+2\nu)$, that involves the Poisson ratio $\nu$.
We thus see that the field $\str$ represents torques and forces (tensions), while $P$ is a linear density of angular and linear momenta.
We observe that both $U$ and $V$ represent elements of the special Euclidean algebra $\mathfrak{se}(3)$, while $P$ and $\str$ are in the dual algebra $\mathfrak{se}(3)^*$ (isomorphic to $\mathfrak{se}(3)$ in this finite-dimensional setting).

To derive the evolution equations for the rod in terms of the evolution of the pairs $(V,U)$ or $(P,\str)$ we start from the variational approach of Holm and Ivanov~\cite{HolIva14} specialized to the quadratic Lagrangian action
\[
\mathcal S=\frac{1}{2}\int_0^{+\infty}\int_0^L\big[V\cdot \mathbb M V-(U-\bar{U})\cdot \mathbb A (U-\bar{U})\big]\,dsdt. 
\]
This involves the total kinetic energy of the rod and the total elastic energy and, taking the first variation of $\mathcal S$, we can apply Hamilton's principle
and obtain the evolution equations for the conservative dynamics of an elastic rod.

It is now important to specify what type of variations are appropriate. 
In fact, the fields $U$ and $V$ as functions of $(s,t)$ are associated with derivatives of the rod placement $\mathcal P$, that is in one-to-one correspondence with a strand in $\mathit{SE}(3)$.
The latter is a time-dependent curve $g:[0,L]\times[0,+\infty)\to \mathit{SE}(3)$ and so we need to consider variations of $U$ and $V$ that are constrained to be consistent with the geometric nature of the rod descriptions.

\subsubsection{Compatibility condition}

The elements of the algebra $\mathfrak{se}(3)$ are tangent vector to $\mathit{SE}(3)$ at the identity.
The strain and spin-velocity matrices at $(s,t)$ are tangent vectors at $g(s,t)\in\mathit{SE}(3)$ generated by motions along $s$ or $t$, respectively, and, in a suitable matrix representation, are given by
\[
\vt L(s,t)=g^{-1}(s,t)\frac{\de g}{\de s}(s,t)\quad\text{and}\qquad \vt\Omega(s,t)=g^{-1}(s,t)\frac{\de g}{\de t}(s,t).
\]
By taking derivatives of the forgoing expressions we obtain
\begin{align*}
\dot{\vt L}&=-g^{-1}\frac{\de g}{\de t}g^{-1}\frac{\de g}{\de s}+g^{-1}\frac{\de^2 g}{\de t\de s}=-\vt\Omega\vt L+g^{-1}\frac{\de^2 g}{\de t\de s},\\
\vt\Omega'&=-g^{-1}\frac{\de g}{\de s}g^{-1}\frac{\de g}{\de t}+g^{-1}\frac{\de^2 g}{\de s\de t}=-\vt L\vt\Omega+g^{-1}\frac{\de^2 g}{\de t\de s}.
\end{align*}
From the difference of these equations and, considering the equality of cross derivatives, we arrive at the compatibility condition
\begin{equation}\label{eq:cc}
\dot{\vt L}-\vt\Omega'=\vt L\vt\Omega-\vt\Omega\vt L.
\end{equation}

\subsubsection{Adjoint operator and its dual}

In the matrix representation of $\mathfrak{se}(3)$ we can identify commutators with adjoint operators as
\[
\ad_{\vt L}(\vt\Omega)=\vt L\vt\Omega-\vt\Omega\vt L=-\ad_{\vt\Omega}(\vt L) .
\]
In the six-component vector representation in which $\vt L\sim U$ and $\vt\Omega\sim V$ we consistently define
\[
\ad_X:=\begin{pmatrix}
0& X_3& -X_2& 0& 0& 0 \\
-X_3& 0& X_1& 0& 0& 0 \\
X_2& -X_1& 0& 0& 0& 0 \\
0& X_6& -X_5& 0& X_3& -X_2 \\
-X_6& 0& X_4& -X_3& 0& X_1 \\
X_5& -X_4& 0& X_2& -X_1& 0 
\end{pmatrix},
\]
where $X_k$ denotes the $k$-th component of the vector $X$.

We also need to compute the dual operator, namely the adjoint-transpose operator $\ad^\tsp$, that is defined in relation to a duality pairing which, in our case, is the Euclidean scalar product in $\R^6$. For all $X, Y\in \mathfrak{se}(3)$ and $Z\in \mathfrak{se}(3)^*\cong \mathfrak{se}(3)$, we have
\[
\ad^\tsp_X(Z)\cdot Y:=Z\cdot\ad_X(Y),
\]
from which we can infer that, in this representation, $\ad^\tsp_X=(\ad_X)^\tsp$.

\subsubsection{Constrained variations}

The Euler--Poincar\'e evolution equations for an $\mathit{SE}(3)$-strand with Lagrangian action $\mathcal S$ can be derived by considering the constrained variations
\[
\delta V=\dot{X}+\ad_VX\quad\text{and}\quad \delta U=X'+\ad_UX
\]
for an arbitrary test field $X:[0,L]\times[0,+\infty)\to \mathfrak{se}(3)$.
In fact, if we consider a variation $\delta g$ of $g$ in the group $\mathit{SE}(3)$ we have $X\sim g^{-1}\delta g$ and
\begin{multline*}
\delta V \sim \delta(g^{-1}\dot{g})=-g^{-1}\delta{g}g^{-1}\dot{g}+g^{-1}\delta\dot{g}+g^{-1}\dot{g}g^{-1}\delta{g}-g^{-1}\dot{g}g^{-1}\delta{g}\\
=(g^{-1}\delta\dot{g}-g^{-1}\dot{g}g^{-1}\delta{g})+g^{-1}\dot{g}g^{-1}\delta{g}-g^{-1}\delta{g}g^{-1}\dot{g})\sim \dot{X}+\ad_VX.
\end{multline*}
An analogous computation proves the expression for $\delta U$. 

By assuming that the variation field $X$ vanishes at the ends of the domain of integration, the first constrained variation of the action $\mathcal S$ gives
\begin{multline}
\langle\delta \mathcal S,X\rangle=\int_0^{+\infty}\int_0^L\big[\delta V\cdot \mathbb M V-\delta U\cdot \mathbb A (U-\bar{U})\big]\,dsdt\\
=\int_0^{+\infty}\int_0^L\big[(\dot{X}+\ad_VX)\cdot P-(X'+\ad_UX)\cdot \str\big]\,dsdt\\
=\int_0^{+\infty}\int_0^L\big[(-\dot{P}+\ad^\tsp_VP)\cdot X-(-\str'+\ad^\tsp_U\str)\cdot X\big]\,dsdt.
\end{multline}
The stationarity condition $\langle\delta \mathcal S,X\rangle=0$ for any test field $X$ then implies
\begin{equation}\label{eq:EP}
\dot{P}-\str'=\ad^\tsp_VP-\ad^\tsp_U\str,
\end{equation}
that encodes the local balances of linear and angular momentum for an elastic rod in the absence of external or dissipative forces.
More general boundary conditions on $X$ can be considered for practical purposes, but the evolution equations would remain the same, and only the boundary conditions satisfied by the solutions would change.
We stress that the derivation of equation \eqref{eq:EP} does not rely upon the choice of a linear constitutive equation.
In fact, it preserves its form if we simply define $\str$ as the derivative of the elastic energy density with respect to $U$.

\subsubsection{Dynamic equations for a viscoelastic rod}

Now that we have derived from a variational principle the conservative evolution equations \eqref{eq:EP}, we are in a position of including additional force and torque densities of a possibly dissipative nature.
The simplest choices of dissipative phenomena consist in an internal dissipation that depends linearly on the time derivative of the strains and an external viscous drag that depends linearly on the generalized velocity.
With these choices, the evolution equations for the dynamics of a viscoelastic rod are
\begin{equation}
\dot{P}-\str'=\ad^\tsp_V P-\ad^\tsp_U \str-\mathbb{D}_\mathrm{ex}V-\mathbb{D}_\mathrm{in}\dot{U}+F,\label{eq:motion0}
\end{equation}
\begin{equation}
\dot{U}-V'=\ad_UV,\label{eq:motion2}
\end{equation}
where the second relation translates the compatibility condition \eqref{eq:cc}, namely the constraint imposed on the rigid-body motion of each cross section by the fact that they should collectively move as a continuous rod.
The six-component vector $F$ represents external force and torque densities acting on each cross section, while the (symmetric positive definite) matrices $\mathbb{D}_\mathrm{ex}$ and $\mathbb{D}_\mathrm{in}$ contain the damping coefficients associated with external drag and internal mechanical dissipation, respectively.

We can substitute definitions \eqref{eq:conjugate_vars} in \eqref{eq:motion0} and \eqref{eq:motion2} and obtain
\begin{equation}
\dot{P}-\str'=\ad^\tsp_{\mathbb M^{-1}P} P-\ad^\tsp_{\mathbb A^{-1}\str+\bar{U}} \str-\mathbb{D}_\mathrm{ex}\mathbb M^{-1}P-\mathbb{D}_\mathrm{in}\mathbb A^{-1}\dot{\str}+F,\label{eq:motion1}
\end{equation}
\begin{equation}\label{eq:motion3}
\dot{\str}-\mathbb A(\mathbb M^{-1}P)'=\mathbb A\ad_{\mathbb A^{-1}\str+\bar{U}}(\mathbb M^{-1}P),
\end{equation}
so that \eqref{eq:motion1} and \eqref{eq:motion3} constitute a system of PDEs in the unknown $(P,\str)$.

\subsection{Discretization}

\begin{figure}
\centering
\includegraphics[width=\textwidth]{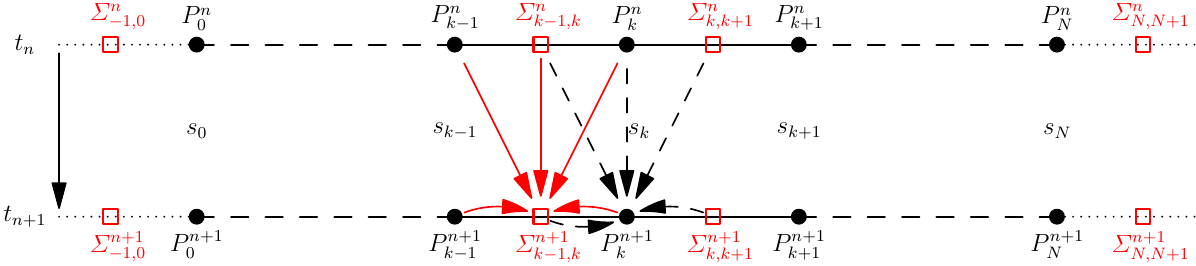}
\caption{The vector $P$ of momenta is represented by values collocated on the nodes of the partition $\{0=s_0,s_1,\ldots,s_N=L\}$ (black circles) while the vector $\str$ of stresses by values collocated on the midpoints of the intervals (red squares). 
Accessory values $\str^n_{-1,0}$ and $\str^n_{N,N+1}$ are collocated outside the domain to properly impose boundary conditions.
In the semi-implicit scheme for the time integration of the evolution, the momenta $P^{n+1}_k$ at time $t_{n+1}$ and point $s_{k}$ are influenced by the values of $\str$ in the neigboring cells at both times $t_n$ and $t_{n+1}$ (black dashed arrows). 
Similarly, the stresses $\str^{n+1}_{k-1,k}$ collocated at $(s_{k-1}+s_k)/2$ are influenced by the values of $P$ on the neighboring nodes at both times $t_n$ and $t_{n+1}$ (red solid arrows).
These local interactions are reflected in the sparse structure of the evolution matrix.}\label{fig:sketch}
\end{figure}

We discretize the evolution equations using a standard semi-implicit time integration that linearizes the evolution operator.
On the other hand, some care is in order when considering the spatial discretization. 
We used finite differences on a staggered grid that mirrors the character of our variables: momenta and velocities are assigned to nodal cross sections, while stresses and strains are viewed as segment quantities and collocated at the midpoint of each mesh interval (Figure~\ref{fig:sketch}).

We introduce a partition $\{0=s_0,s_1,\ldots,s_N=L\}$ of $[0,L]$ and consider a space-time cell for $s_k\leq s\leq s_{k+1}$ and $t^n\leq t\leq t^{n+1}$ and denote by a subscript $k$ or $k,k+1$ nodal or segment quantities, respectively. Superscripts indicate the time instant at which the quantity is computed.
Boundary conditions are imposed by adding accessory cells and nodes at the two ends of the rod. 
In this way, it is rather straightforward to drive or fix the motion of the rod ends or to set free-end conditions.
Within this scheme, equation \eqref{eq:motion1} features descretized quantities that live on the nodes $s_k$ of the partition, while \eqref{eq:motion3} features descretized quantities that live on the intervals of the partition.

The discretization of equation \eqref{eq:motion1} reads, for inner and free nodes with $0<k\leq N$,
\begin{multline}
\frac{P^{n+1}_{k}-P^{n}_{k}}{t^{n+1}-t^n}-\frac{\str^{m}_{k,k+1}-\str^{m}_{k-1,k}}{(s_{k+1}-s_{k-1})/2}=\mathrm{ad}^\tsp_{V^n_{k}} P^{m}_{k}-\frac{1}{2}[\mathrm{ad}^\tsp_{U^n_{k-1,k}} \str^{m}_{k-1,k}+\mathrm{ad}^\tsp_{U^n_{k,k+1}} \str^{m}_{k,k+1}]\\
-\mathbb{D}_\mathrm{ex}\mathbb M^{-1}P^{m}_{k}-\mathbb{D}_\mathrm{in}\mathbb A^{-1}\frac{\str^{n+1}_{k,k+1}-\str^{n}_{k,k+1}}{t^{n+1}-t^n}+F^{n}_{k},
\end{multline}
and the discretization of equation \eqref{eq:motion3} for inner segments ($0<k<N-1$) is
\begin{multline}
\frac{\str^{n+1}_{k,k+1}-\str^{n}_{k,k+1}}{t^{n+1}-t^n}-\mathbb A\frac{\mathbb M^{-1}P^m_{k+1}-\mathbb M^{-1}P^m_k}{s_{k+1}-s_k}=
\frac{1}{4}\mathbb A\ad_{U^n_{k,k+1}}\big(\mathbb M^{-1}P^m_{k}+\mathbb M^{-1}P^m_{k+1}\big)\\
-\frac{1}{4}\mathbb A\ad_{(V^n_{k}+V^n_{k+1})}\big(\mathbb A^{-1}\str^m_{k,k+1}{+\bar{U}_{k,k+1}}\big),
\end{multline}
with the substitutions $U^n_{k,k+1}=\mathbb A^{-1}\str^n_{k,k+1}+\bar{U}_{k,k+1}$ and $V^n_{k}=\mathbb M^{-1}P^n_{k}$.
The choice $m=n$ describes a fully explicit scheme, while $m=n+1$ leads to a semi-implicit scheme, that becomes fully implicit only in those cases in which the nonlinear terms are exactly vanishing (purely axial, shearing or twisting deformations; bending is excluded).

Different boundary conditions can be imposed but, for the following tests, we always need the same set of conditions.
At one end of the rod we prescribe the motion through a given $\bar{P}_0(t)$, possibly vanishing, while $\str$ is free; on the other end we have a given stress $\str=\bar{\str}_\mathrm{e}(t)$ and $P$ free. These conditions translate into
\begin{gather}
P^{n+1}_{0}=\bar{P}_0(t),\\
\str^{n+1}_{-1,0}-\str^{n+1}_{0,1}=0,\\
\str^{n+1}_{N-1,N}+\str^{n+1}_{N,N+1}=2\bar{\str}_\mathrm{e}(t),
\end{gather}
where pedices $(-1,0)$ and $(N,N+1)$ denote the accessory segments that are added outside the physical rod to impose the boundary conditions.

\section{Numerical results}\label{sec:examples}

To assess the effectiveness of our method we present a series of examples and some results about computational costs. In the small-displacement regime we can make comparisons with analytical results derived from the linearized equations of motion. On the other hand, in the nonlinear large-displacement regime we will test our numerical solutions against published results on some benchmark problems.
We implemented the computational model in the Python language, exploiting the scientific computing libraries \texttt{numpy} and \texttt{scipy} and the just-in-time compilation features provided by \texttt{numba}.
We tested our implementation on a Laptop with a 1,8 GHz Intel\textsuperscript{\textregistered} Core{\texttrademark} i5 processor and 8 GB of 1600 MHz DDR3 memory.

\subsection{Small-displacement regime: cantilever}

\begin{table}
\centering
\begin{tabular}{r|l}
Parameter & Value\\
\hline
total relaxed length & 4 m \\
inner diameter & 0.1155 m \\
outer diameter & 0.1397 m \\
linear mass density & 34.2277 kg/m \\
Young modulus & 200 GPa \\
Poisson ratio & 0 \\
\hline
\end{tabular}
\caption{Material parameters for the static cantilever experiment.}\label{tab:params_cantilever}
\end{table}

\begin{figure}
\centering
\includegraphics[width=\textwidth]{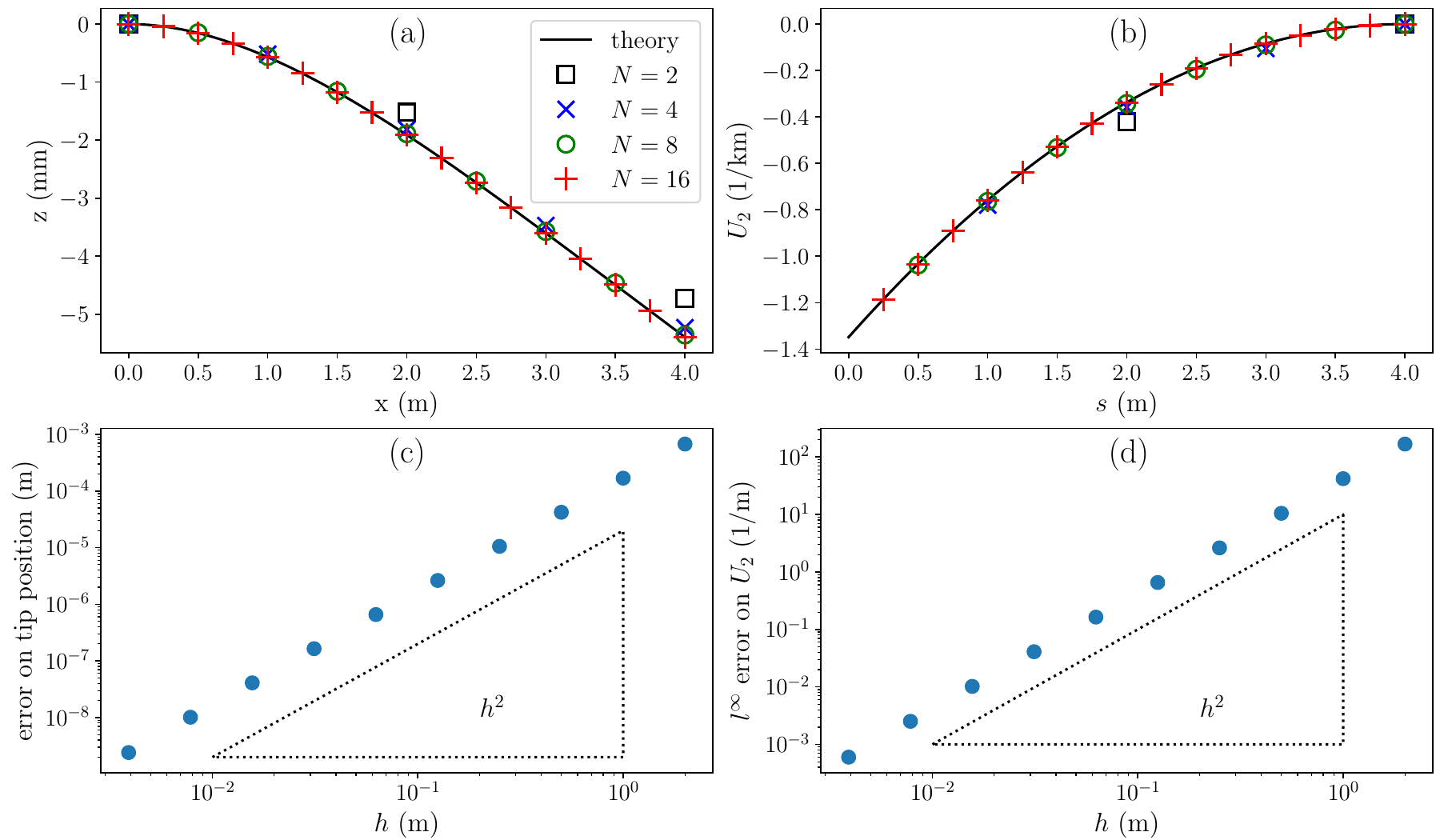}
\caption{The numerical solution for a 4-meter-long cantilever clamped at one end and subject to its own weight (with material parameters given in Table~\ref{tab:params_cantilever}) captures very well the analytic curves for the midline displacement (a) and curvature (b) already with a small number $N$ of segments. The error on the tip displacement (c) and on the curvature $U_2$ (d) scale as $h^{2}$, $h=L/N$ being the size of the mesh intervals.}\label{fig:cantilever_static}
\end{figure}

We compared the static solution for a cantilever, clamped at one end and subject to its own weight, obtained as a long-time limit of the dissipative dynamics for a 4 m long hollow cylinder with material parameters given in Table~\ref{tab:params_cantilever}.
The dissipation, useful to reach the static solution, is generated by the internal dissipation matrix $\mathbb D_\mathrm{in}=\eta_\mathrm{in}\mathrm{diag}(1,1,1,1,1,1)$, with $\eta_\mathrm{in}=10^{-4}$.
The spatial domain is discretized uniformly with a variable number $N$ of intervals.
With a time step of $0.01$ s, we obtain convergence to a static solution (identified by a kinetic energy below $10^{-12}$ J) within 100 steps in all cases.
The results, presented in Figure~\ref{fig:cantilever_static}(a,b), show a very good approximation of the analytical solution already with $N=8$.
To assess the order of convergence of our approximation, we computed the $l^\infty$-norm of the difference between the reconstructed nodal values of $U_2$ for each $N$ and the numerical solution for $N=4096$, namely $\sup_{s\in[0,L]}|U_2^{(N)}(s)-U_2^{(4096)}(s)|$. 
We found that this error estimate scales as $h^{2}$, where $h=L/N$ is the size of the mesh intervals.
If we consider only the linear terms of the evolution equations, this result is consistent with our discretization that employs centered finite differences.
The same scaling can be observed for the absolute error on the tip displacement (Figure~\ref{fig:cantilever_static}(c,d)).

We then analyzed the vibration of a cantilever with the same physical parameters given above, initialized with a small curvature ($0.01$ m$^{-1}$) and in the absence of gravity and dissipation. We varied the length but kept the number of discrete segments equal to 16. We compared the fundamental frequency of the tip displacement given by the theory, $\nu_\mathrm{theory}$, with those computed from our solution by means of a Fast Fourier Transform algorithm, $\nu_\mathrm{fft}$. The results, presented in Table~\ref{tab:frequency}, show a very good match.

\begin{table}
\begin{center}
\begin{tabular}{r|ccccc}
length (m) & 1 & 2 & 4 & 8 & 16 \\
\hline
$\nu_\mathrm{theory}$ (Hz) & 135.1 & 33.8 & 8.44 & 2.11 & 0.528 \\
$\nu_\mathrm{fft}$ (Hz) & 133.1 & 33.6 & 8.43 & 2.11 & 0.528 \\
\hline
\end{tabular}
\end{center}
\caption{Comparison of  the fundamental frequency of the tip displacement under vibrations of cantilevers with different length as given by the theory and computed from our numerical solution.}\label{tab:frequency}
\end{table}

\subsection{Large-displacement regime: static solution}

\begin{figure}
\centering
\includegraphics[width=\textwidth]{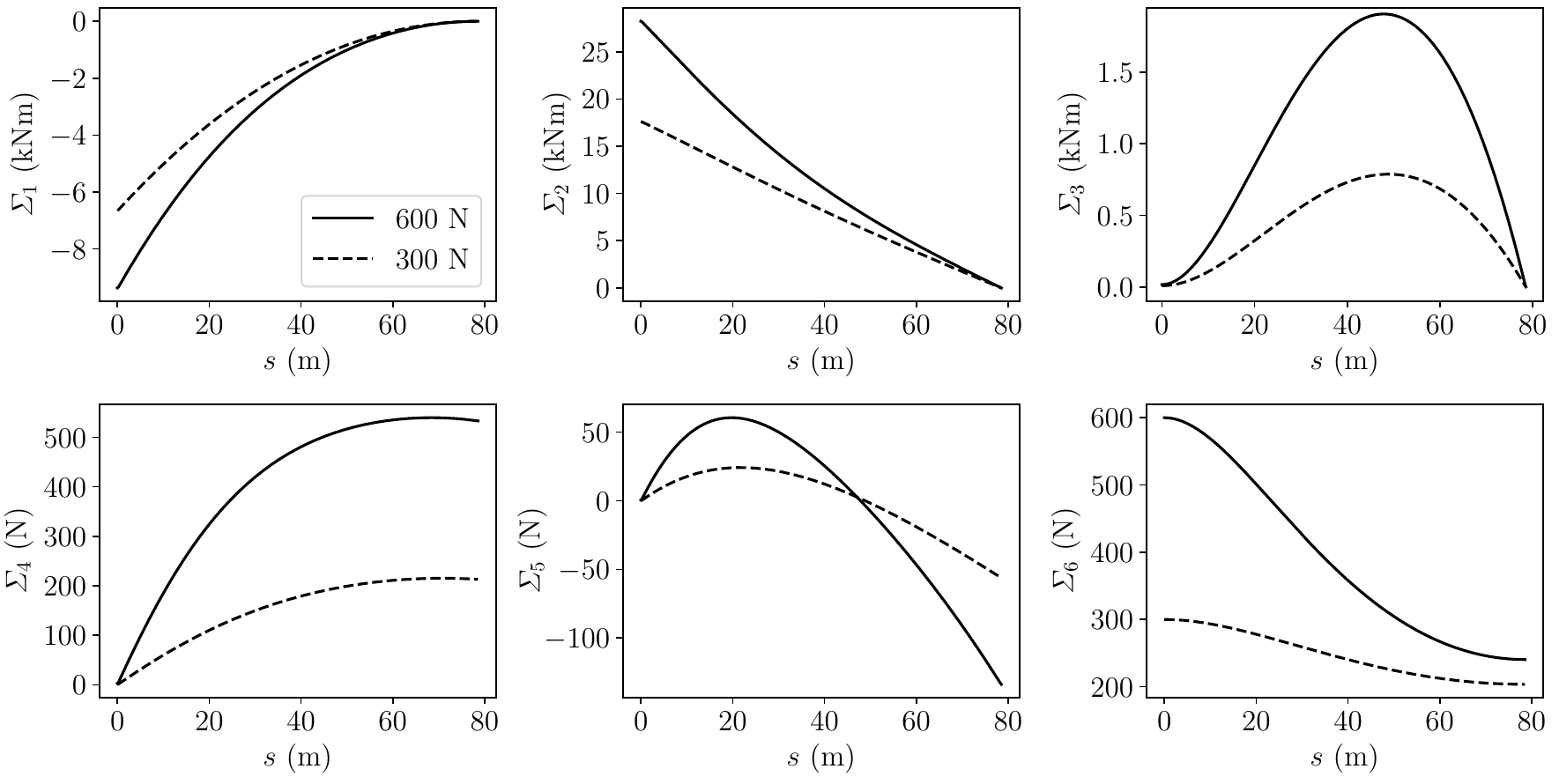}
\caption{The vertical load at the tip of the 45-degree bend activates all of the components of the generalizad stress $\str$. At the tip $s=L$, the first three components, associated with twisting ($\str_1$) and bending ($\str_2$, $\str_3$) moments, vanish, while those associated with stretching ($\str_4$) and bending ($\str_5$, $\str_6$) are given by the projection of the imposed load on the local directors $(\vc d_1(L),\vc d_2(L),\vc d_3(L))$. The reported data are obtained with a mesh of $N=320$ segments.}\label{fig:45-stress}
\end{figure}

\begin{figure}
\centering
\includegraphics[width=\textwidth]{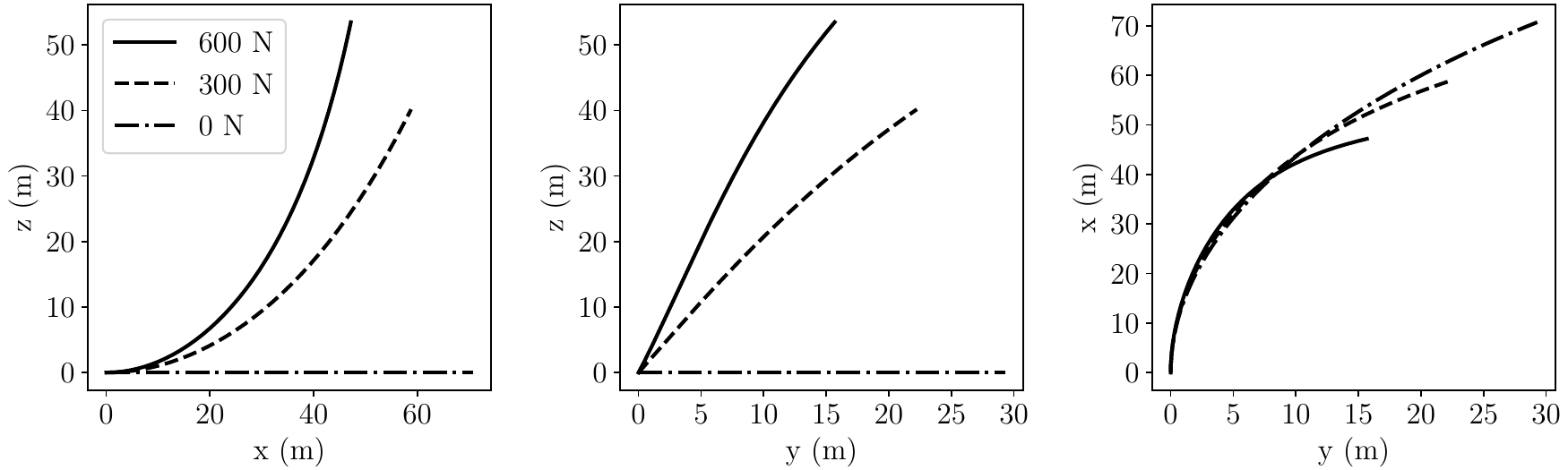}
\caption{The deformation of the planar relaxed configuration (dot-dashed curve) induced by the applied vertical loads at the tip brings the rod into a three-dimensional deformation that activates the geometric nonlinearities of the stress response. We report projections on the coordinate planes of the midline $\vc x(s)$ for $s\in[0,L]$, with $\vc x(0)$ at the origin and $\vc x(L)$ at the other end.}\label{fig:45-pos}
\end{figure}

To validate our model in the large-displacement regime, in which all of the strains are activated and coupled by the geometric nonlinearities, we reproduced a common test case and compare our results with those reported by Simo \& Vu-Quoc~\cite{SimVuQ86}, Sonneville, Cardona \& Br\"uls~\cite{SonCar14} and Howcroft \emph{et al.}~\cite{HowCoo18}.
For further comparison of the computational performance, we solved the same problem with the commercial software Abaqus\textsuperscript{\textregistered}, using quadratic beam elements \texttt{B32}.
The problem consists in a cantilever 45-degree bend subjected to a fixed load at the tip (the effect of weight is neglected).
The relaxed configuration (Figure~\ref{fig:45-pos}, dot-dashed curves) spans a circular arc of 45 degrees in the $xy$-plane.
The beam has a square cross section of side 1 m, radius of curvature of 100 m and total length $L=78.54$ m. The Young modulus is $E=10^7$ Pa and the Poisson ratio $\nu=0$.
The rod is initially in the horizontal plane and the load is applied in the vertical direction.
The equilibrium configuration is computed for two different values of the load (300 N and 600 N).
Due to the nonlinear coupling between all of the strains produced by the curved geometry of the beam, all of the components of $\str$ are activated (Figure~\ref{fig:45-stress}) and contribute to determine the equilibrium configuration (Figure~\ref{fig:45-pos}).
The good agreement between our method and those presented in the literature can be assessed by considering the tip displacement reported in Table~\ref{tab:tip}, where, for an easier comparison of the computational efficiency, we consider a discretization with 81 nodes.
As shown in Table~\ref{tab:times}, while the number of degrees of freedom we use is comparatively large (as typical of local discretization schemes) the computation remains very fast.

\begin{table}
\begin{center}
\begin{tabular}{r|lll|lll}
load (N) & 300 &  &  & 600 & & \\
\hline
displacement (m) & x & y & z & x & y & z \\
\hline
Simo \& Vu-Quoc~\cite{SimVuQ86} & 58.84 & 22.33 & 40.08 & 47.23 & 15.79 & 53.37 \\
Sonneville \emph{et al.}~\cite{SonCar14} & 58.84 & 22.30 & 40.03 & 47.23 & 15.76 & 53.28 \\
Howcroft \emph{et al.}~\cite{HowCoo18} & --- & --- & --- & 46.90 & 15.55 & 53.60 \\
Abaqus\textsuperscript{\textregistered} & 58.42 & 21.97 & 40.29 & 45.83 & 15.46 & 53.37 \\
Present method & 58.86 & 22.23 & 40.11 & 47.25 & 15.64 & 53.43 \\ 
\hline
\end{tabular}
\end{center}
\caption{Comparison of tip displacement data for the 45-degree bend example.}\label{tab:tip}
\end{table}

\begin{table}
\begin{center}
\begin{tabular}{r|l|l|l}
 & discretization & DOF & time (s) \\
\hline
Howcroft \emph{et al.}~\cite{HowCoo18} & 11 shapes & 11 & 3.6 \\
MSC Nastran\textsuperscript{\textregistered}~\cite{HowCoo18} & 22 elements & 132 & 40.1 \\
Intrinsic beam~\cite{HowCoo18} & 31 elements & 186 & 132 \\
Abaqus\textsuperscript{\textregistered} & 150 elements & 1800 & 4 \\ 
Present method & 81 nodes & 960 & 0.8 \\ 
\hline
\end{tabular}
\end{center}
\caption{Comparison of degrees of freedom (DOF) and computation time for the 45-degree bend with a load of 600 N.}\label{tab:times}
\end{table}

\begin{figure}[h]
\centering
\includegraphics[width=\textwidth]{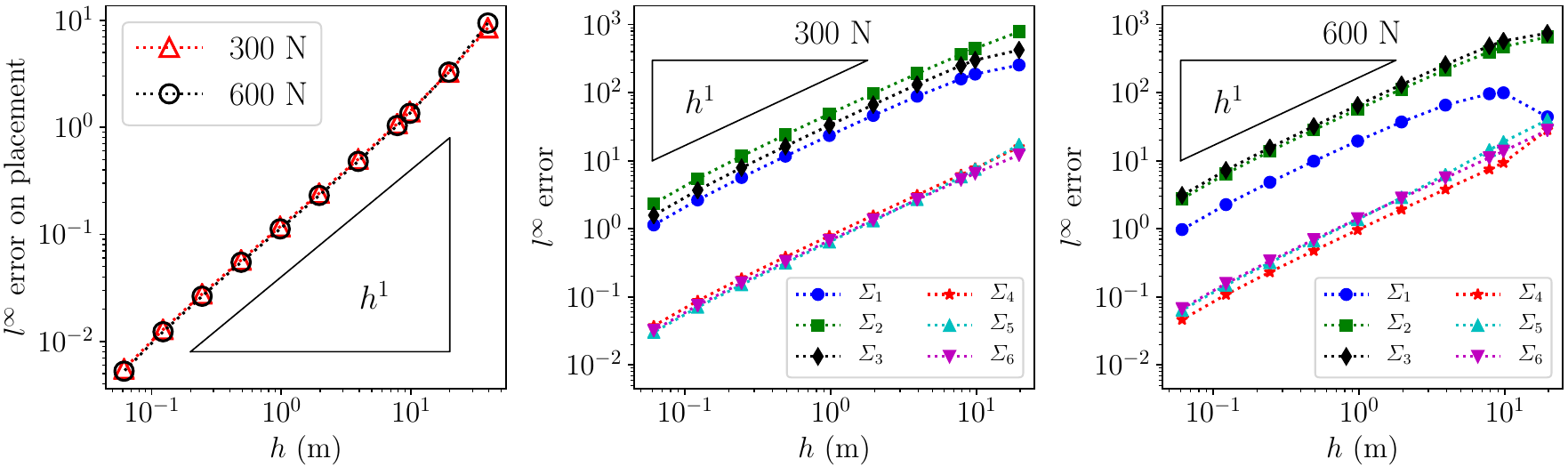}
\caption{Scaling of error estimates on the numerical solution with respect to the mesh size $h=L/N$ for the 45-degree bend test. The solution with $N=5120$ segments is taken as reference. The $l^\infty$ error on the midline placement scales linearly in $h$ and so does the $l^\infty$ error on the components of $\str$, the actual degrees of freedom we solve for in our scheme. 
The difference in comparison to the small-displacement regime can be attributed to the different weight of the nonlinear terms in the solution, that entails an approximation of order $h^2$ for terms that are quadratic in $\str$.}\label{fig:scaling}
\end{figure}

We studied the convergence of the numerical approximation by computing solutions for different values of $N$, using up to $N=5120$ segments.
Similar to what was done for the cantilever, the static solution is achieved following a dissipative dynamics.
In this case, we added an external dissipation matrix $\mathbb D_\mathrm{ex}=\eta_\mathrm{ex}\mathrm{diag}(1,1,1,1,1,1)$, with $\eta_\mathrm{ex}=10^{-1}$.
With a time step of $10$ s the solution converges within $100$ steps for each value of $N$.
The $l^\infty$ error on the three-dimensional rod configuration is computed as the maximum distance between images of the same abscissa (mesh node) through the midline placement, namely $\sup_{s\in[0,L]}|\vc{x}^{(N)}(s)-\vc{x}^{(5120)}(s)|$. It turns out to corresponds to the error on the tip position and to be linear in $h=L/N$.
Similarly, the $l^\infty$ error on the solution for each component of $\str$, that is $\sup_{s\in[0,L]}|\str_i^{(N)}(s)-\str_i^{(5120)}(s)|$ for $i=1\ldots,6$, scales as $h$ (Figure~\ref{fig:scaling}).
The difference from the quadratic scaling observed in the small-displacement regime can be attributed to the different weight of the nonlinear terms in the solution, that entails an approximation of order $h^2$ for terms that are quadratic in $\str$.

\subsection{Large-displacement regime: dynamic solution}

\begin{figure}
\centering
\includegraphics[width=\textwidth]{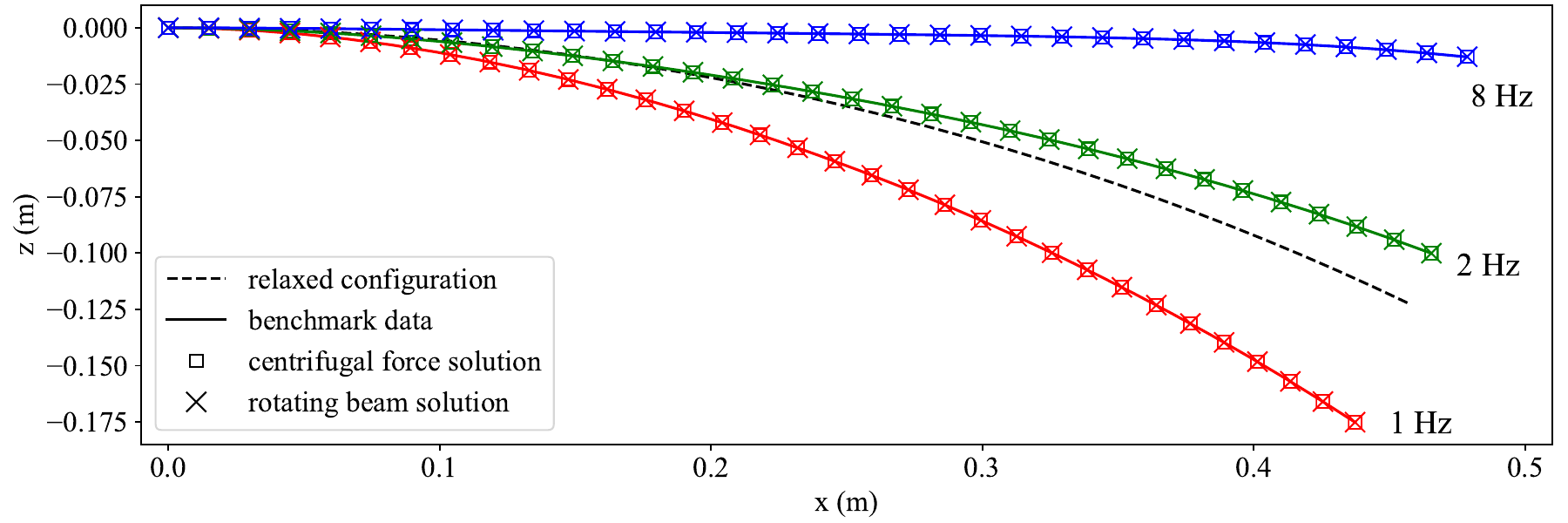}
\caption{We validate our method in the nonlinear dynamic regime by comparing with the data reported by Howcroft \emph{et al.}~\cite{HowCoo18} (solid lines) the radial profile of the midline for a beam (material parameters given in Table~\ref{tab:blade}) with a curved relaxed configuration (dashed line) that is spinning around the vertical axis at different frequencies (1 Hz, 2 Hz, and 8 Hz). The symbols represent solutions obtained with $N=32$ segments. We check the internal consistency of our method by superimposing the profile generated by a rotating beam (crosses) with the static solution in a rotating frame with centrifugal forces (squares).}\label{fig:rotating_steady}
\end{figure}

The geometric nonlinearities that characterize the rod dynamics in the large-displacement regime have a strong influence on both the transient and steady motion behavior of slender structures. To provide a first check that our method is able to correctly capture such nonlinear effects we use as benchmark the data published by Howcroft \emph{et al.}~\cite{HowCoo18} about two tests. We consider a rather flat cantilever beam with material parameters given in Table~\ref{tab:blade}. One end of the beam is clamped so that the tangent to the midline at $s=0$ points always in the $xy$-plane. The motion of that end is driven as detailed below. The stress-free configuration features an intrinsic curvature $\bar{U}_2=-3\pi/(18L)$ m$^{-1}$ that points the tip (at $s=L$) a little downward (Figure~\ref{fig:rotating_steady}, dashed curve). The dynamics is dissipative, with only internal dissipation as in the first example with $\eta_\mathrm{in}=10^{-1}$, taken from the benchmark case. 

\begin{table}[h]
\centering
\begin{tabular}{r|l}
Parameter & Value\\
\hline
relaxed length & 0.479 m \\
width & $5.08\times10^{-2}$ m \\
height & $4.5\times10^{-4}$ m \\
linear mass density & 0.1012698 kg/m \\
Young modulus & 127 GPa \\
Poisson ratio & 0 \\
\hline
\end{tabular}
\caption{Material parameters for the rotating beam experiment.}\label{tab:blade}
\end{table}

In the first test, we consider the profile of the beam in steady rotational motion around the vertical axis.
This is obtained by imposing at the clamped end a non-vanishing component of the angular momentum along the vertical axis with various frequencies.
The steady radial profiles obtained with $N=32$ segments match very well both the benchmark data and the profile obtained by solving the static problem in a rotating frame, in which we impose the appropriate centrifugal forces (Figure~\ref{fig:rotating_steady}).

\begin{figure}
\centering
\includegraphics[width=\textwidth]{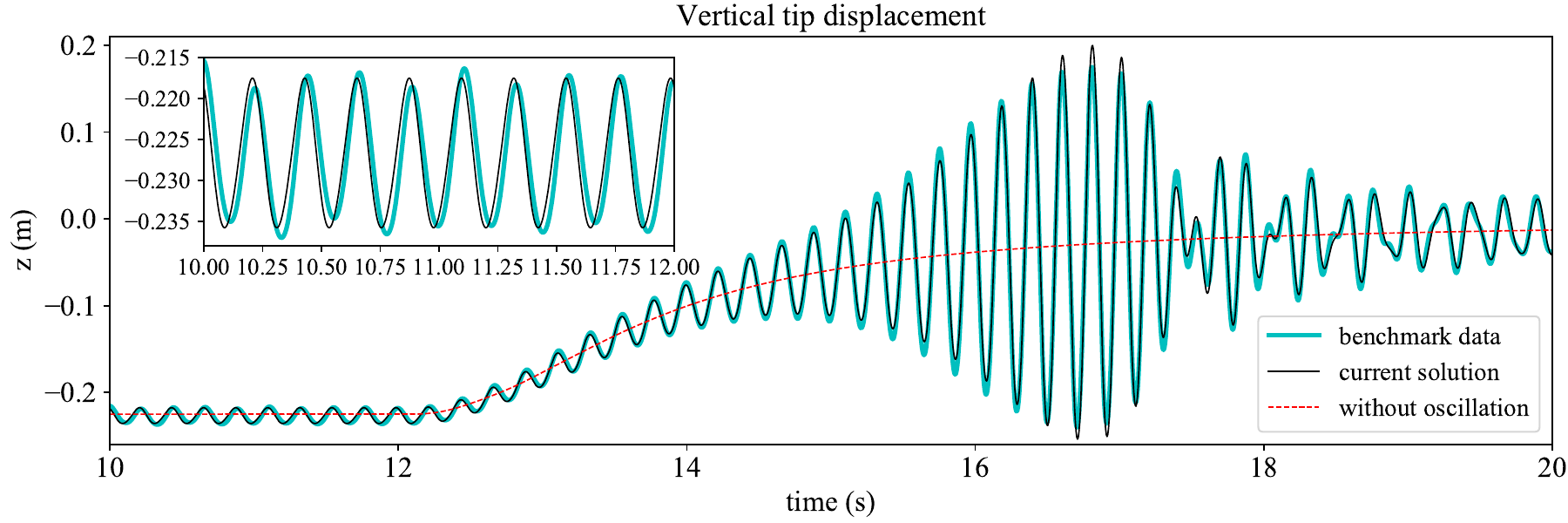}
\caption{The time evolution of the vertical tip displacement (thin solid line) for the curved beam under combined rotation and vertical oscillation of the clamped end is compared with the data reported by Howcroft \emph{et al.}~\cite{HowCoo18} (thicker turquoise line). Up to time $t=12$ s only the vertical oscillation at $4.5$ Hz is active. At that time, the rotation is turned on with a frequency that increases linearly and reaches $8$ Hz at $t=20$ s. The tip oscillates always around the position obtained without the vertical oscillation (dashed red line) but the amplitude of the oscillation presents a resonance phenomenon when an intrinsic vibrational frequency of the nonlinear system, that varies due to the change in curvature, crosses $4.5$ Hz.
Small discrepancies between our data and the benchmark may be attributed to the residual presence in the latter of some oscillation mode different from the $4.5$ Hz signal, probably due to the propagation of transient effects from the initialization of the test that, on the contrary, are completely gone in our data, as can be appreciated from the inset.}\label{fig:rotating_ramp}
\end{figure}

In the second test, the clamped end oscillates vertically with a frequency of $4.5$ Hz and a total amplitude of $0.04$ m. After $12$ s we superimpose to this oscillation a rotation about the vertical axis with a frequency that increases linearly for $8$ s up to $8$ Hz. During this ramp, the stiffening induced by the change in curvature causes an intrinsic vibrational frequency of the nonlinear system to cross $4.5$ Hz, so that a resonance phenomenon can be observed. The time evolution computed with our method nicely captures oscillations and this transient phenomenon, matching very well the selected benchmark data (Figure~\ref{fig:rotating_ramp}).

\subsection{Computational efficiency}

\begin{figure}
\centering
\includegraphics[width=0.5\textwidth]{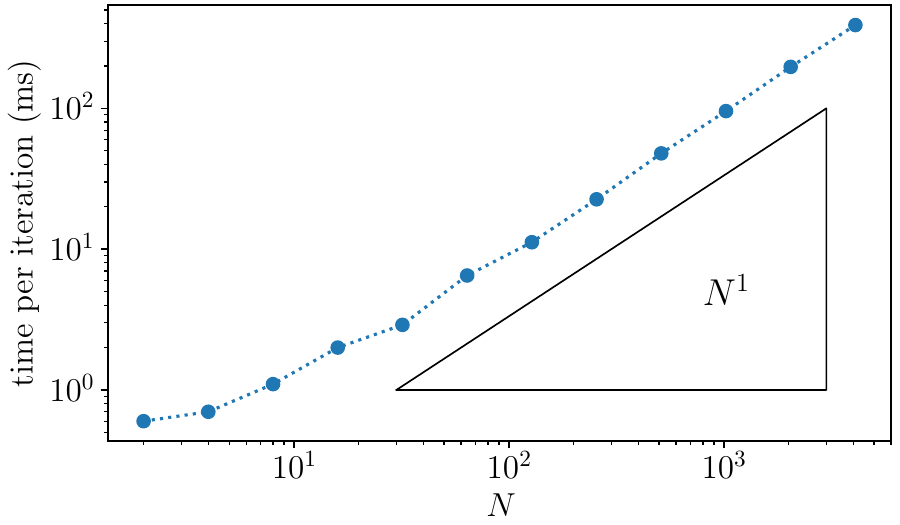}
\caption{The computation time per iteration scales linearly with the number $N$ of segments. The data refer to the solution of the static cantilever performed on a Laptop with a 1,8 GHz Intel\textsuperscript{\textregistered} Core{\texttrademark} i5 processor and 8 GB of 1600 MHz DDR3 memory.}\label{fig:efficiency}
\end{figure}

The semi-implicit computational scheme requires the solution of a linear system with a matrix that is updated at each time step.
The system matrix is banded with 47 non-vanishing diagonals.
The current implementation exploits such a structure in the system solution.
The construction of the matrix is indeed the most expensive part of the algorithm and we achieved an optimal memory usage and a computation time that scales linearly with the number $N$ of mesh intervals, proportional to the degrees of freedom (Figure~\ref{fig:efficiency}).

\section{Conclusions}\label{sec:conclusion}

We presented a method for the simulation of the dynamics of slender structures described within the framework of special Cosserat rods.
The beam is viewed as a one-parameter family of rigid cross sections. 
For this reason, the Lie group $\mathit{SE}(3)$ of rigid-body motions and the associated Lie algebra $\mathfrak{se}(3)$ play a fundamental role in the description of the rod kinematics and dynamics.
In fact, one can identify the placement of the rod in the three-dimensional ambient space as an $\mathit{SE}(3)$-strand, namely a curve in $\mathit{SE}(3)$, and deduce the intrinsic evolution equations for an elastic rod from a variational principle, computing the Euler--Poincar\'e equations associated with an $\mathit{SE}(3)$-invariant Lagrangian action.

As is well known from the literature, the system Lagrangian only involves elements of the algebra $\mathfrak{se}(3)$ that describe both the generalized velocity of each cross section and the generalized strains of the rod.
One usually needs to take into account kinematic relations between the two that involve translational and rotational degrees of freedom, but this geometric setting produces additional compatibility equations, involving solely generalized velocities and strains, that encode the constraint imposed on the rigid-body motion of each cross section by the fact that they should collectively move as a continuous rod.

We extend the evolution equations to incorporate dissipative effects that can be of internal origin, depending on variations of the strains over time, and external, such as the viscous drag exerted by a surrounding fluid.
These terms do not require additional degrees of freedom to be expressed and are thus perfectly compatible with the proposed intrinsic framework.
In our presentation, we focused on linear constitutive prescription for both the elastic and the viscous response, but the inclusion of nonlinear laws is mathematically trivial, though it may be computationally more challenging.
It should be noted that the linear elastic laws allow for a formulation of the equations that is simple and minimal, in the sense that we have twelve equations for twelve unknown fields, whereas the treatment of nonlinear constitutive laws may be more manageable in a mixed formulation involving eighteen equations, six of which would be the algebraic constitutive relations.

The viscoelastic evolution equations together with the consistency relations constitute a theoretically sound and practically flexible starting point for numerical approximation schemes.
We have implemented a finite-difference approximation on a staggered grid in space and a semi-implicit time-stepping that, in spite of its simplicity, shows a very good computational performance in paradigmatic tests for both static and dynamic problems.
Moreover, its good scalability makes it particularly attractive for the treatment of very large structures.

Within our general setting one can impose internal constraints such as unshearbility and inextensibility with a minimal effort, since they translate into linear constraints on the set of strains.
Moreover, the fast reconstruction of the rod placement, that can be performed by applying the exponential map (see \ref{app:a}) to the generalized velocity at each time step, allows for the inclusion of position-dependent external forces that may be of relevance for the simulation of contact and other external interactions.

\section*{Acknowledgments}

This research has been supported by Tenaris (\texttt{www.tenaris.com}).

\appendix

\section{Exponential map in the Special euclidean setting}\label{app:a}

To reconstruct the final placement $\mathcal P_\mathrm{f}$ of a cross section that, starting from a given $\mathcal P_\mathrm{i}$, moves for a small time interval $\tau$ with constant spin--velocity matrix $\vt\Omega$ can be achieved by means of the matrix exponential function, that maps the element $\tau\vt\Omega\in\mathfrak{se}(3)$ into an element of the matrix representation of the group $\mathit{SE}(3)$, namely
\[
\mathcal P_\mathrm{f}=\exp(\tau\vt\Omega)\mathcal P_\mathrm{i}.
\]
It is well known that the accurate computation of a generic matrix exponential may be challenging, but thanks to the peculiar structure of matrices that represent elements of $\mathfrak{se}(3)$, we can easily find that
\begin{equation}\label{eq:expm}
\exp(\tau\vt\Omega)=\mathsf{Id}+\tau\vt\Omega+\frac{1-\cos\theta}{\theta^2}(\tau\vt\Omega)^2+\frac{\theta-\sin\theta}{\theta^3}(\tau\vt\Omega)^3\,,
\end{equation}
where $\theta=\sqrt{\tau^2(w_1^2+w_2^2+w_3^2)}$ is the norm of the spin component of $\tau\vt\Omega$.
With this analytical relation we can keep track of the rod placement as a post-processing of the solution to the evolution equations, that may also be important in calculating position-dependent forces acting on the system.
It is important to observe that the expression \eqref{eq:expm} is ill-conditioned for $\theta\to 0$ and we found it convenient to replace it with its Taylor expansion up to $o(\theta^6)$ for $\theta<0.1$.

A completely analogous formula allows to reconstruct the placement of the rod starting from $\mathcal P_0$ at one end an iteratively computing
\[
\mathcal P_{k+1}=\exp\big((s_{k+1}-s_k)\vt L_{k,k+1}\big)\mathcal P_{k},
\]
where $\vt L_{k,k+1}$ is the matrix associated with the strains assumed constant on the segment $[s_k,s_{k+1}]$.

\section*{Data Availability Statement}

The data that support the findings of this study are available from the corresponding author upon reasonable request.



\end{document}